\documentclass{emulateapj}

\slugcomment{ApJ, in press}
\shorttitle{Magnetic Massive Stars in NGC 3766}
\shortauthors{McSwain}

\begin{document}

\title{Detection of Magnetic Massive Stars in the Open Cluster NGC 3766}

\author{M.\ Virginia McSwain\altaffilmark{1}}
\affil{Department of Physics, Lehigh University, Bethlehem, PA 18015}
\email{mcswain@lehigh.edu}

\altaffiltext{1}{Visiting Astronomer, Paranal Observatory}


\begin{abstract}
A growing number of observations indicate that magnetic fields are present among a small fraction of massive O- and B-type stars, yet the origin of these fields remains unclear.  Here we present the results of a VLT/FORS1 spectropolarimetric survey of 15 B-type members of the open cluster NGC 3766.  We have detected two magnetic B stars in the cluster, including one with a large field of nearly 2 kG, and we find marginal detections of two additional stars.  There is no correlation between the observed longitudinal field strengths and the projected rotational velocity, suggesting that a dynamo origin for the fields is unlikely.  We also use the Oblique Dipole Rotator model to simulate populations of magnetic stars with uniform or slightly varying magnetic flux on the ZAMS.  None of the models successfully reproduces our observed range in $B_\ell$ and the expected number of field detections, and we rule out a purely fossil origin for the observed fields.  
\end{abstract}


\section{Introduction}

Magnetic fields have long been inferred in solar and late-type stars and are produced by a dynamo mechanism in the outer convection zones.  Studies of hot star structure and evolution have long neglected magnetic fields since the radiative envelopes of hot stars are not expected to produce them.  Yet, an increasing number of observations have revealed that magnetic fields are present in $\sim 5$\% of hot stars \citep{petit2008}, even though the origin of such fields is uncertain \citep{neiner2007}.  Various models have proposed that they may be generated in the convective cores, although no known mechanism can transport them to the surface in a timescale consistent with observations \citep{charbonneau2001, macgregor2003}.  Other models have proposed that the radiative envelopes somehow maintain a dynamo process near the stellar surface (e.g.\ \citealt{spruit2002}; \citealt{macdonald2004}).  Finally, magnetic fields in hot stars may be a relic from stellar formation, hence the present day field is a ``fossil'' of the primordial field. 

A growing number of studies have begun to investigate magnetic fields among massive stars in open clusters.  Main-sequence (MS) and pre-main sequence (PMS) magnetic OB stars have recently been detected in the Orion Nebula Cluster \citep{petit2008}, NGC 6611, and NGC 2244 \citep{alecian2008} with a variety of ages, rotational periods, and chemical abundances.  Currently, the most widescale investigation of magnetic, early-type cluster members considers 258 A- and B-type stars in more than 40 open clusters \citep{bagnulo2006, landstreet2008}.  These studies offer strong potential to infer the origin of the magnetic fields in massive stars.  If the magnetic fields are relics of the molecular cloud from which the stars formed, then a population of coeval OB stars should offer widespread evidence of the initial field.  A past or present dynamo likely depends on other physical conditions within the stars, especially rotation, and would likely be less widespread among cluster members.  

Observations currently suggest a fossil origin for the magnetic fields in massive stars.  \citet{landstreet2008} find that the strongest fields are associated with stars near the zero-age main-sequence (ZAMS) and that the field strength decays over time. For all but the strongest fields, the decay is consistent with magnetic flux conservation as the stars increase in radius during the MS lifetime.  \citet{landstreet2008} also find that some of the strongest fields are found in the most slowly rotating stars, and they find no correlation between rotation and magnetic field strength as expected for a magnetic dynamo in hot stars.  However, there is no evidence that widespread magnetic braking contributes to the loss of angular momentum in B-type stars.  The rotational velocity of B stars has been observed to decline with age \citep{huang2006}, but no more than expected for their increase in radius during the MS lifetime and angular momentum losses due to stellar winds.

The open cluster NGC 3766 is between 14.5 and 25 Myr in age (WEBDA; \citealt{moitinho1997}; \citealt{tadross2001}), with a reddening $E(B-V) = 0.22 \pm 0.03$ and at a distance of 1.9 to 2.3 kpc \citep{mcswain2008}.  In our spectroscopic study of this cluster \citep{mcswain2008}, we measured projected rotational velocities, $V \sin i$, for 37 members.  We also measured the effective temperatures, $T_{\rm eff}$, and polar surface gravities, $\log g_{\rm polar}$, for 42 cluster members.  More recently, we have analyzed spectra of additional cluster members using the same technique, and these will be presented in a forthcoming work (McSwain et al., in prep).  In these works, we have identified several B-type members with unusually high or low helium abundances that are excellent candidates for the presence of magnetic fields.  Since NGC 3766 offers a number of likely magnetic field candidates as well as a chance to study the rotational and evolutionary dependence of hot star magnetic fields, it is an excellent target to search for magnetic B stars.

For these reasons, we have conducted a spectropolarimetric survey of 15 B-type stars in NGC 3766.  In this work, we present the definite detection of two magnetic stars in the cluster, including one strong field of almost 2 kG.  We also identify two stars with marginal magnetic field detections that are worth further investigation.  The observations and magnetic field measurements are discussed in Section 2.  In Section 3, we discuss the possibility that the stellar magnetic field strength is correlated with rotation and/or MS evolution.  We also present a method to test whether the observed magnetic field distribution is consistent with a fossil field origin.  Our conclusions are summarized in Section 4.


\section{Observations and Data Reduction}

We observed 15 members of NGC 3766 during two nights of visitor mode time at Paranal Observatory on UT dates 2008 March 24--25.  Our targets were selected to include as many helium strong and helium weak stars as possible, as well as Be stars and normal B-type stars in the cluster.  They span a large range of $V \sin i$ and include both MS and giant members.  We did not include any supergiants since we expect that magnetic fields are expected to be stronger in stars with higher $\log g_{\rm polar}$ due to the conservation of magnetic flux.
We used the FORS1 instrument, mounted on the 8m \textit{Kueyen} (Unit 2) telescope of the \textit{VLT}, in spectropolarimetric mode with the super-achromatic quarter-wave phase retarder plate in front of a Wollaston prism with a beam divergence of 22'' in standard resolution mode.  We used GRISM 600B, with 600 grooves~mm$^{-1}$ and a slit width of 0.4'', to achieve a spectral resolving power $R =1500-2500$.  The spectra cover the wavelength range 3250--6180 \AA, which includes all hydrogen Balmer lines except H$\alpha$.  

In order to measure magnetic fields with a 100 G formal error bar with this instrumental setup, a total photon count per pixel of about $10^5$ is necessary \citep{bagnulo2002}.  Therefore we used a standard readout mode with low gain (A, $1\times1$, low) and exposure times between 200--900 s to achieve a typical signal of 10,000--30,000 photons per pixel in each exposure.  To improve the signal, we obtained a continuous series of six exposures, alternating the retarder waveplate at two different angles ($+45^\circ$ and $-45^\circ$).  Because of the long exposure times needed, we were only able to observe cluster members brighter than $V < 11$.  Each target was observed at the center of the field of view to avoid off-axis, spurious instrumental polarization signals \citep{bagnulo2002}.  Each exposure simultaneously recorded the ordinary and extraordinary beams of the Wollaston prism on separate regions of the chip.  Table \ref{observations} lists a summary of the observations.
 
\placetable{observations}

Dome flats were taken during the day with the telescope pointed at the zenith and using both retarder waveplate setups used during the observations.  Bias frames and helium lamp wavelength calibration images were also taken each day. The spectra were bias subtracted, flat fielded, cosmic ray cleaned, and wavelength calibrated in IRAF using standard reduction procedures for slit spectroscopy.

Since the wavelength calibrations were taken during the day, the spectra for each night are on an identical wavelength grid and no interpolation was necessary.  For each star, we obtained Stokes $I$, the intensity in the unpolarized spectrum, by summing over all 12 available beams.  The $I/I_c$ spectrum was rectified to a unit continuum using line free regions.  We calculated the $V$ spectrum using the method advocated by \citet{bagnulo2002},
\begin{equation}
\frac{V}{I} = \frac{1}{2} \left [ \left (\frac{f^o - f^e}{f^o + f^e} \right )_{\alpha=-45^\circ} - \left (\frac{f^o - f^e}{f^o + f^e} \right )_{\alpha=+45^\circ} \right ]
\end{equation}
to minimize cross-talk effects and imperfect flat-fielding correction.  Here, $\alpha$ gives the position angle of the retarder waveplate and $f^o$ and $f^e$ are the ordinary and extraordinary beams, respectively.  

\begin{figure}
\includegraphics[angle=90,scale=.24]{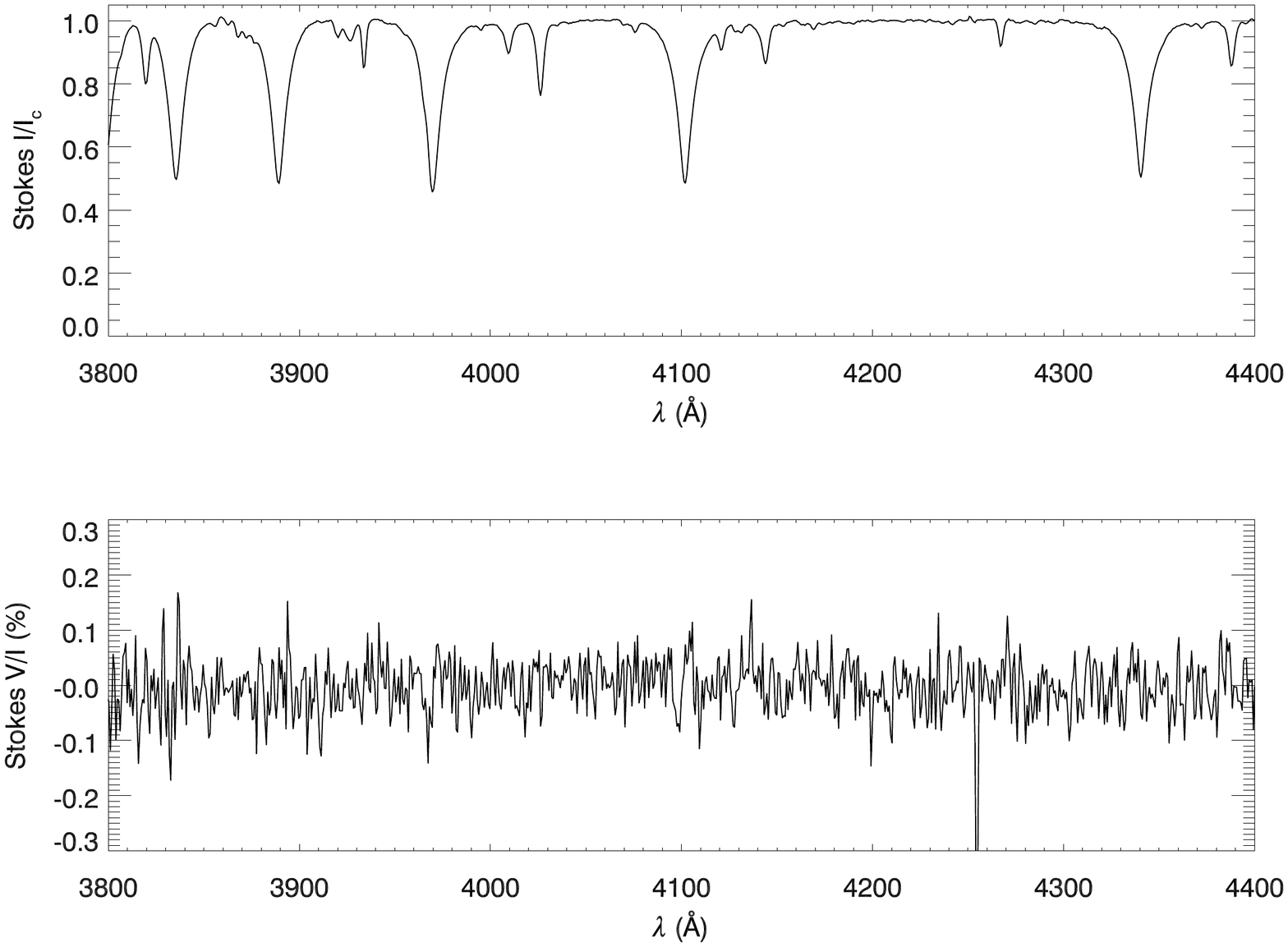}
\includegraphics[angle=90,scale=.24]{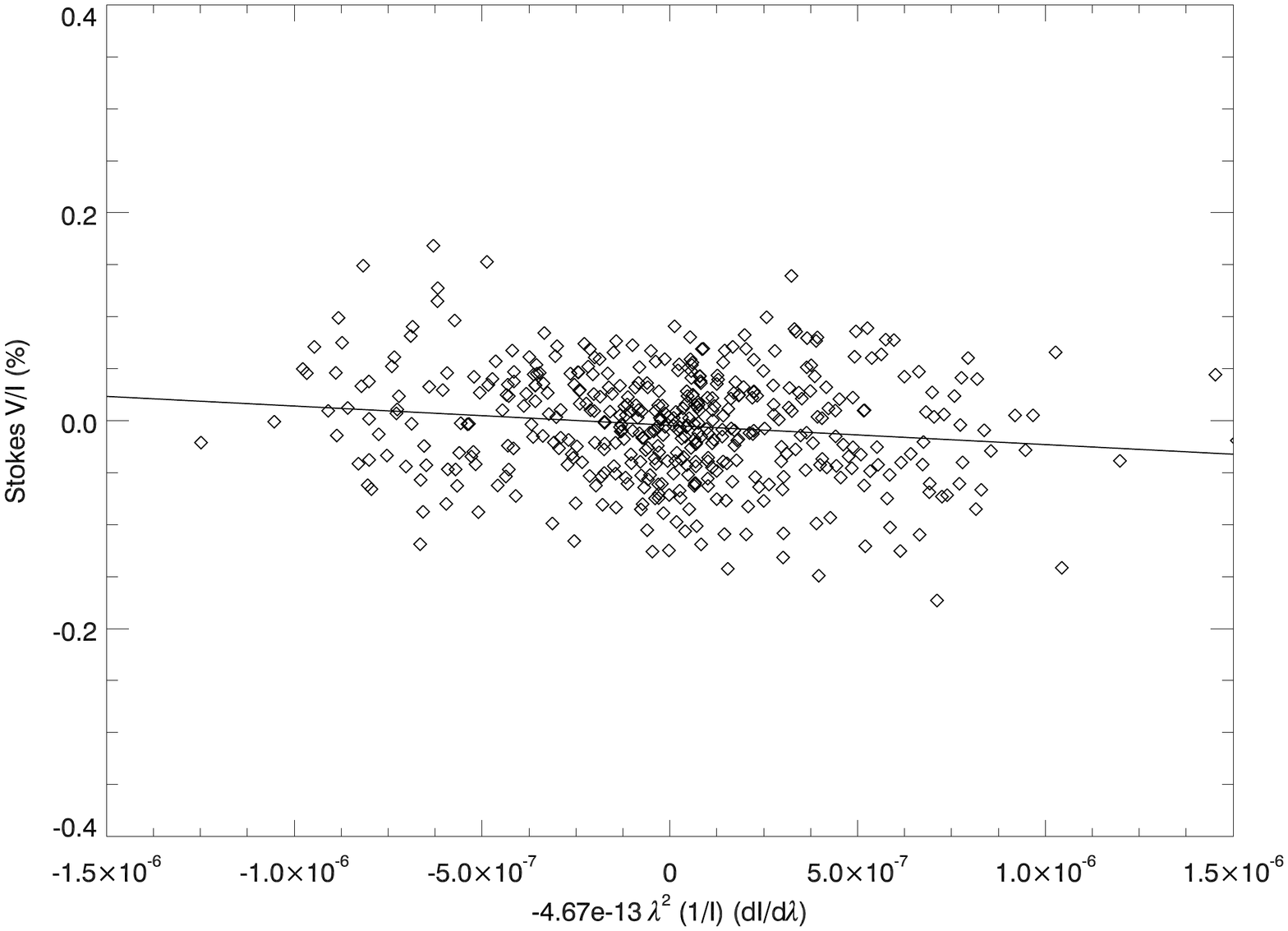}
\caption{(left) Stokes $I/I_c$ and $V/I$ spectra of No.\ 45.   (right)  $B_\ell$ is proportional to the slope of the least-square linear fit to the observed data.  We find a marginal detection of $B_\ell = -185 \pm 53$ G for No.\ 45.
\label{star45}
} 
\end{figure}

\begin{figure}
\includegraphics[angle=90,scale=.24]{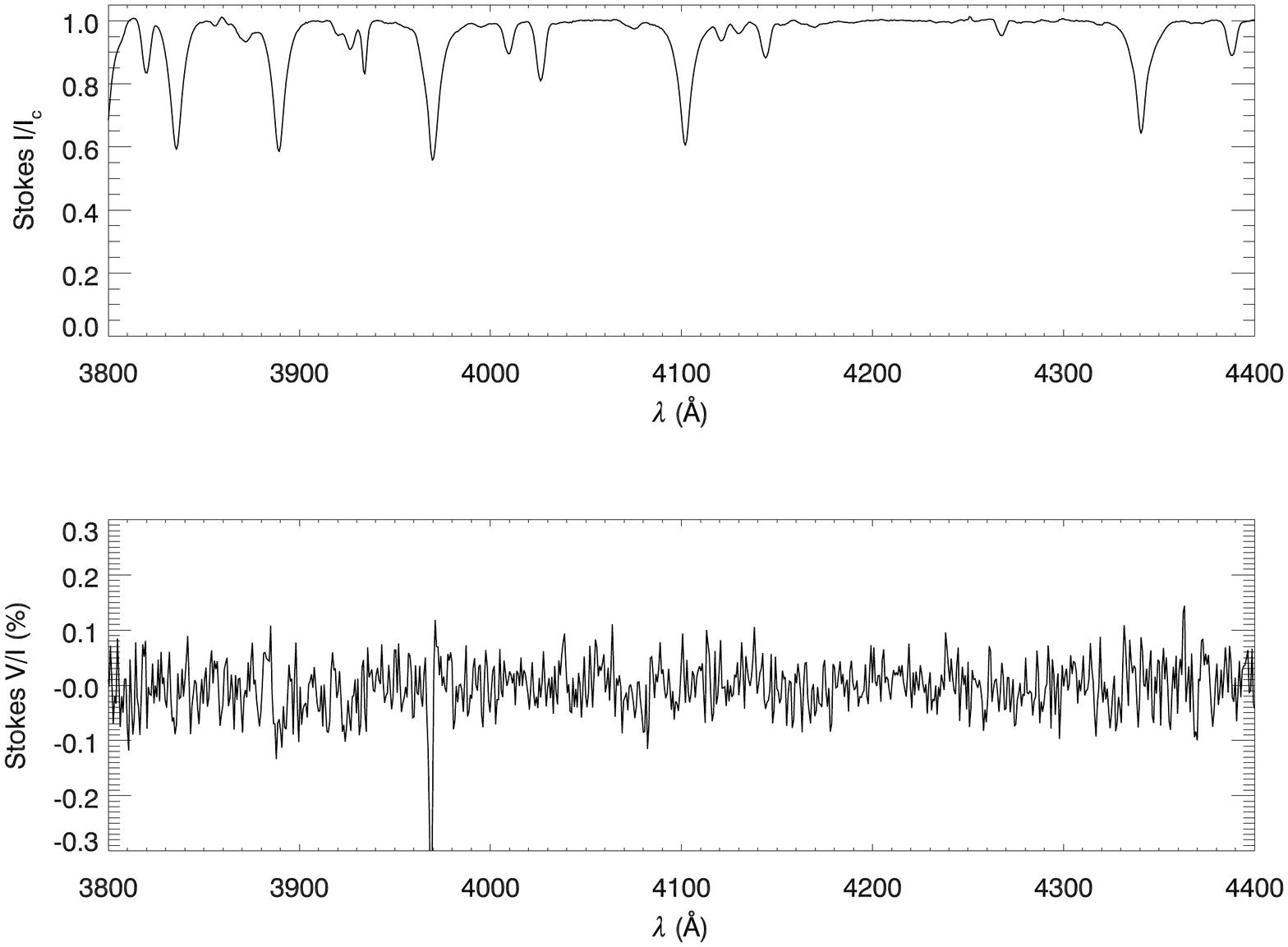}
\includegraphics[angle=90,scale=.24]{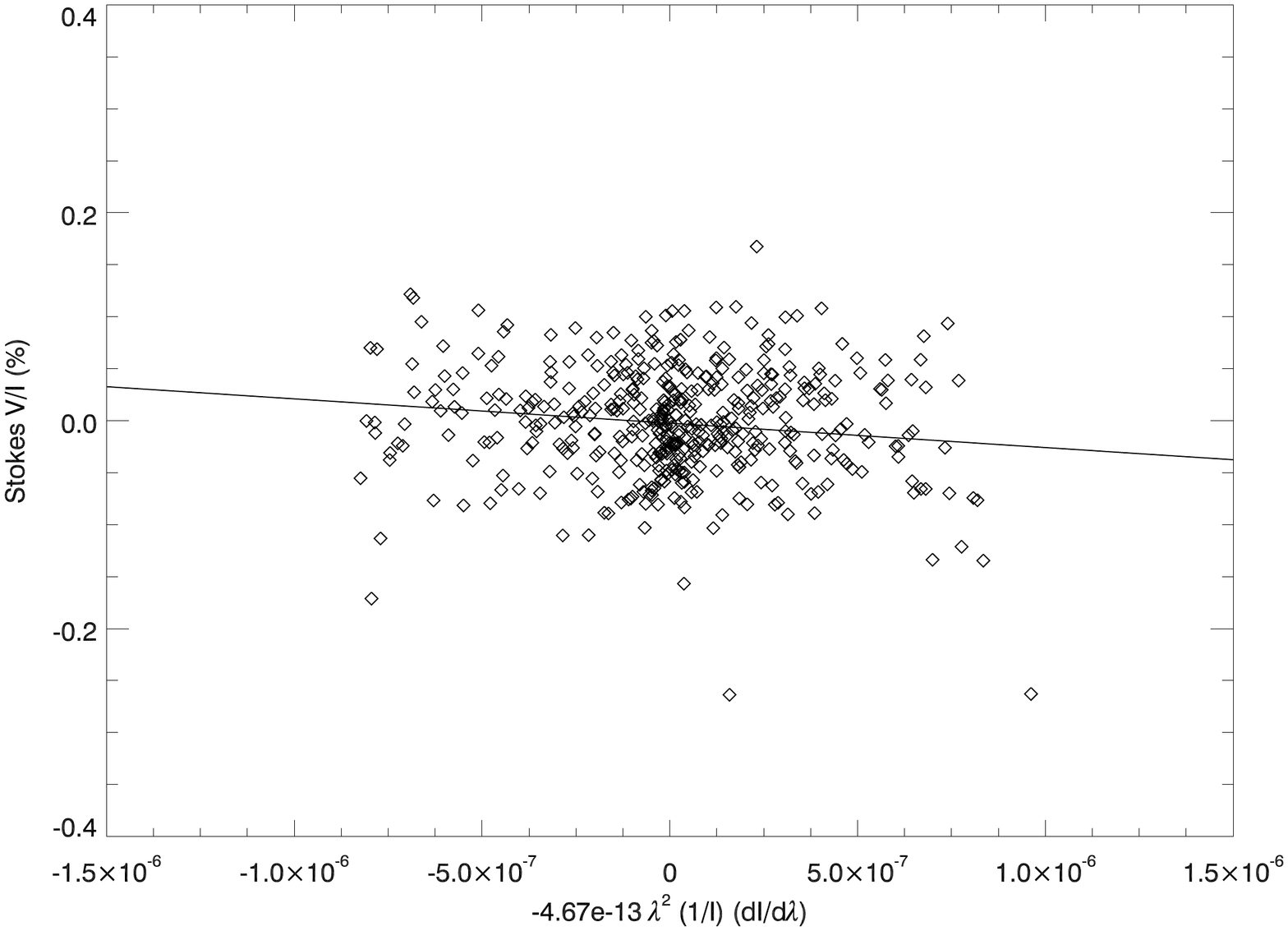}
\caption{(left) Stokes $I/I_c$ and $V/I$ spectra of No.\ 47.   (right)  $B_\ell$ is proportional to the slope of the least-square linear fit to the observed data.  We find a marginal detection of $B_\ell = -234 \pm 69$ G for No.\ 47.
\label{star47}
} 
\end{figure}

\begin{figure}
\includegraphics[angle=90,scale=.24]{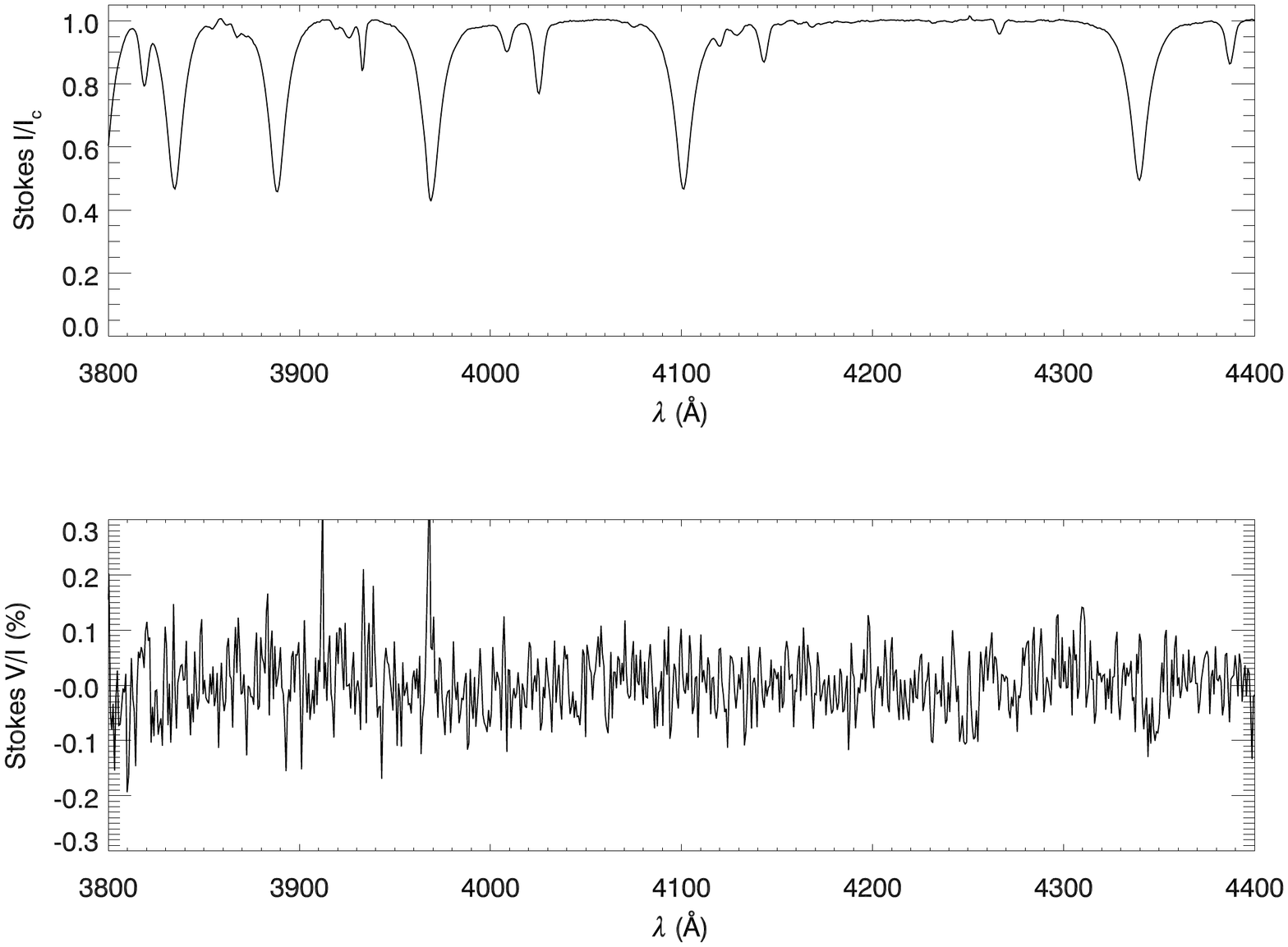}
\includegraphics[angle=90,scale=.24]{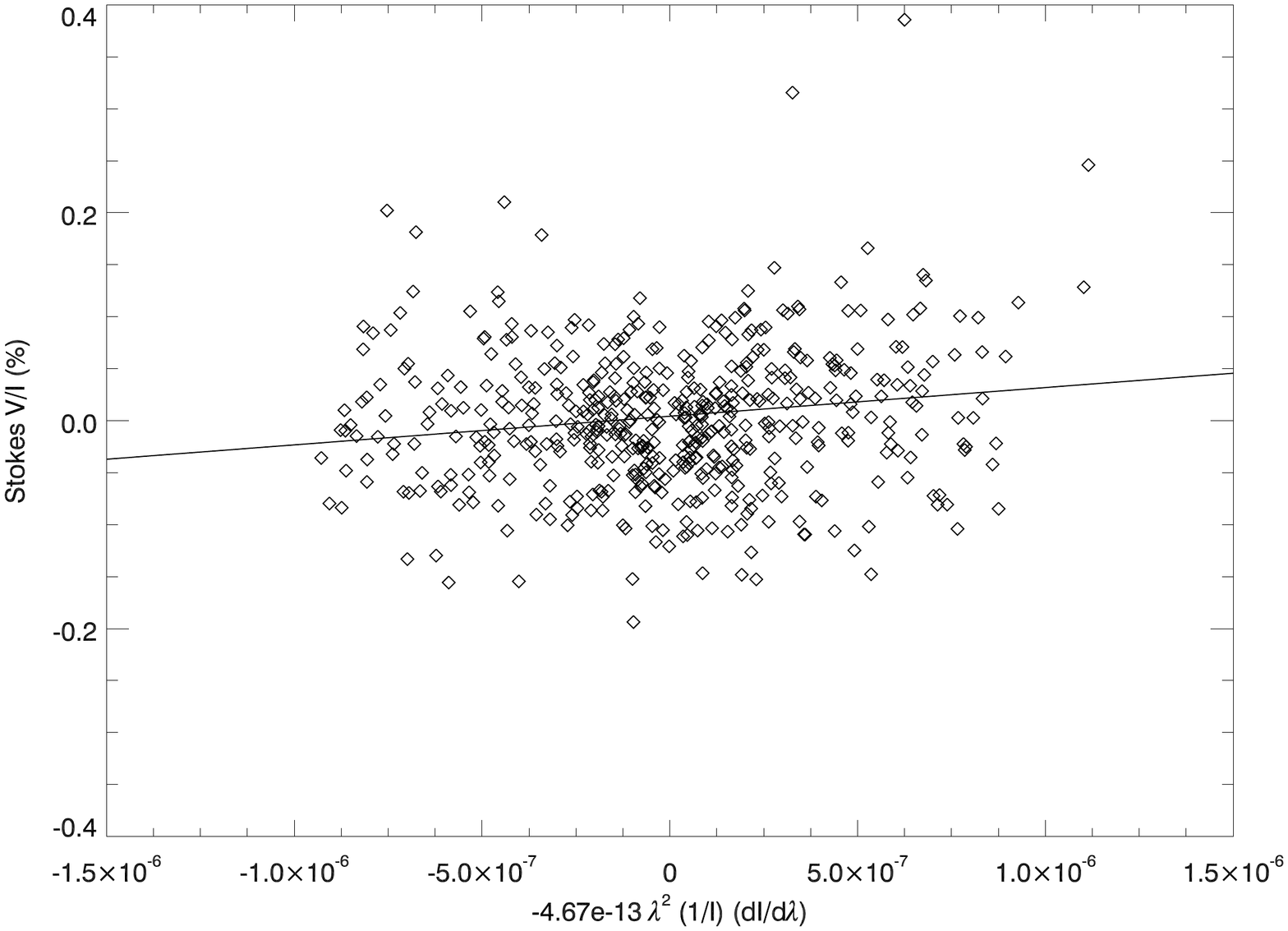}
\caption{(left) Stokes $I/I_c$ and $V/I$ spectra of No.\ 94.   (right)  $B_\ell$ is proportional to the slope of the least-square linear fit to the observed data.  We find $B_\ell = 276 \pm 55$ G for No.\ 94.
\label{star94}
} 
\end{figure}

\begin{figure}
\includegraphics[angle=90,scale=.24]{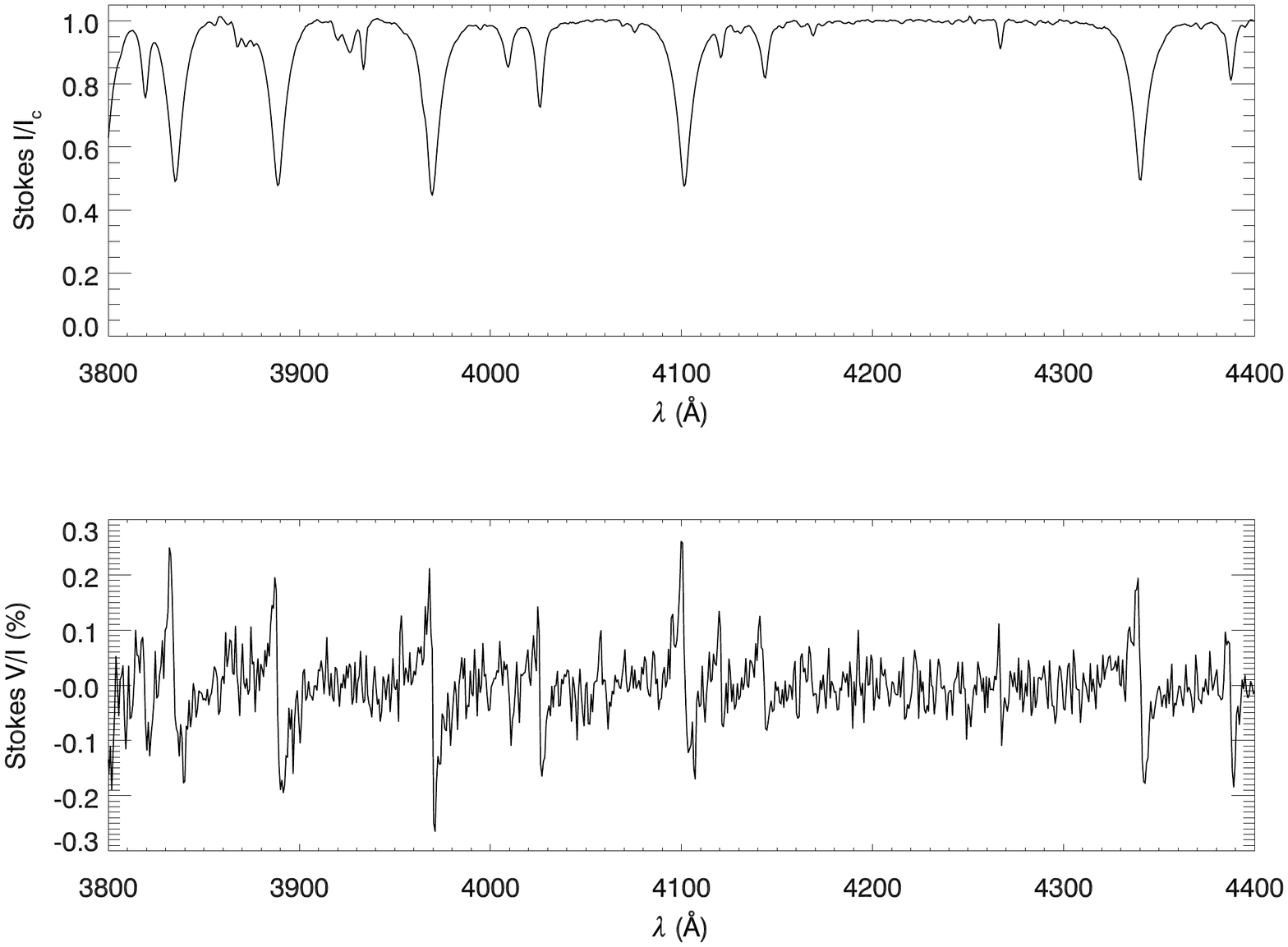}
\includegraphics[angle=90,scale=.24]{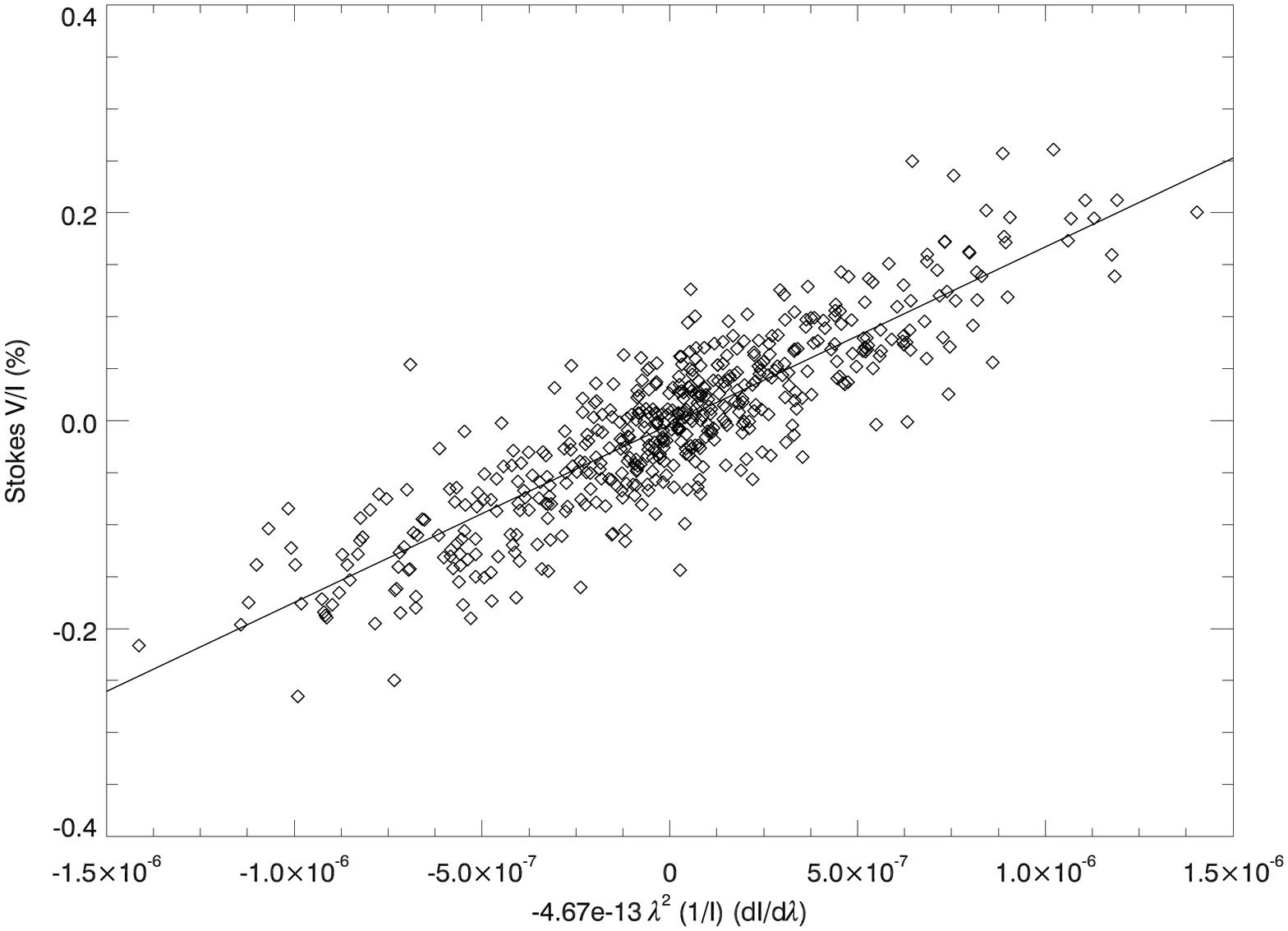}
\caption{(left) Stokes $I/I_c$ and $V/I$ spectra of No.\ 170.   (right)  $B_\ell$ is proportional to the slope of the least-square linear fit to the observed data.  We find $B_\ell = 1710 \pm 32$ G for No.\ 170.
\label{star170}
} 
\end{figure}

\citet{landstreet1982} showed that in the weak field regime, the longitudinal magnetic field, $B_\ell$, can be derived from  the difference between the circular polarizations observed in the red and blue wings of the hydrogen line profiles using 
\begin{equation}
\frac{V}{I} = - \frac{g_{\rm eff} e \lambda^2}{4 \pi m_e c^2} \frac{1}{I} \frac{dI}{d\lambda} B_\ell,
\end{equation}
where $g_{\rm eff}$ is the effective Land\'e factor, $e$ is the electron charge, $\lambda$ is the wavelength, $m_e$ is the electron mass, and $c$ is the speed of light.  \citet{mathys1988} shows that this relationship holds when rotation and limb darkening are taken into account.  In our calculations, we set $g_{\rm eff} = 1$ for the Balmer lines \citep{casini1994} and $g_{\rm eff} = 1.2$ for \ion{He}{1} and metal lines \citep{bagnulo2002}.  Although we did not observe strong emission in the Balmer lines of any Be stars in the sample, we excluded the H$\beta$ line for the Be stars since that line profile is most likely to be contaminated by emission.  We computed the measurement errors of the $V/I$ spectra using the standard deviation of the spectra over line free regions.  We applied this error in $V/I$ to all points of the measured lines and used a least-squares linear fit to measure $B_\ell$ and its error, $\sigma$, according to Equation 2.  We claim a definite magnetic field detection if $|B_\ell|$ is at least $4\sigma$, and marginal if between 2--4$\sigma$.  The resulting measurements are included in Table \ref{observations}.  
For each marginal and definite detection, we show a plot of each Stokes $I/I_c$ and $V/I$ spectra as well as the $B_\ell$ fit in Figures \ref{star45}--\ref{star170}.  No.\ 170 has a strong magnetic field, thus there is a clear polarization signature in the $V/I$ spectrum.  The other detections have somewhat weaker $B_\ell$, so the polarization signature in those $V/I$ spectra is difficult to identify given the lower S/N of those spectra.  Our mean $\sigma$ from our observations is 45 G, implying that on average, our survey has a $4\sigma$ detection limit of $B_{min} = 180$~G.  Even with our worst case $\sigma$, our detection limit is 292 G.


\section{Discussion}

In order to probe a variety of massive star environments, our targets in NGC 3766 reflect many different types of B stars in the cluster.  We provide a summary of each target's physical properties in Table \ref{targets}.  In columns 1--2, we provide the identification numbers from \citet{mcswain2005b} as well as the WEBDA database.  We also give the Str\"omgren $y$ magnitude  from \citet{mcswain2005b} in column 3.  Columns 4--6 provide a physical description of each star ($V \sin i$, $T_{\rm eff}$, $\log g_{\rm polar}$) from our spectroscopic investigations (\citealt{mcswain2008}; McSwain et al.\ in prep).  We used $T_{\rm eff}$ and $\log g_{\rm polar}$ to interpolate between the evolutionary tracks of \citet{schaller1992} to measure each star's mass,  $M_\star$, and ZAMS and current radii, $R_0$ and $R_\star$, respectively.  They are listed in columns 7--9.  Finally, column 10 provides a short comment about the nature of each star.

\placetable{targets}

Since we observed cluster members with a large range in $V \sin i$ and $\log g_{\rm polar}$, it is worth investigating possible trends with those parameters.  We do not observe any trends with the detection of $B_\ell$ with $V \sin i$, suggesting that rotation is not a major factor influencing the presence of magnetic fields.   In fact, star 170 has a strong field of nearly 2 kG, yet it is a relatively slow rotator among B-type stars.  In our small sample, we also do not observe any trends with $\log g_{\rm polar}$, an important indicator of evolution along the MS.  All of our targets are early B-type stars with rather small range in $T_{\rm eff}$ and $M_\star$, so we cannot consider any possible trends with those parameters.  

A number of studies have proposed that B stars with unusual helium abundances are likely to have magnetic fields (see the discussion of \citealt{huang2006}).  Most of the cluster members with abnormally weak \ion{He}{1} lines were too faint to observe ($y > 12$), but the one He weak star observed, No.\ 170, was found to have a very strong magnetic field of 1.7 kG.  We also observed 3 He strong members whose \ion{He}{1} line strength could be a signature of either magnetic fields or blended lines in a binary.  We found a definite detection of a magnetic field in the He strong star 94.  
Our target list also included 5 transient Be stars whose disks have been observed to disappear and/or reappear \citep{mcswain2008} and 3 Be stars with stable disks to investigate whether magnetic behavior might influence the formation of their disks.  Among the Be stars observed, we found marginal detections of $B_\ell$ in No.\ 47.  Finally, we included a control sample of B-type stars with normal abundance patterns and without evidence of the Be phenomenon.  We found a marginal detection in No.\ 45, which we classify as a normal B star since it has not exhibited Balmer line emission in any of our data.  However, it was identified as a possible Be star by \citet{shobbrook1985, shobbrook1987} and may be a transient Be star.  

To look for possible variations in the \ion{He}{1} line profiles that might be due to magnetic activity altering the surface He abundances, especially among the He weak and He strong members, we compared the Stokes $I/I_c$ spectrum of each star to prior spectra obtained with the CTIO 4.0m Blanco telescope (\citealt{mcswain2008}, McSwain et al.\ in prep.).  Nearly all the stars have deeper and sharper lines in our 2006 and 2007 data sets, a consequence of the higher spectral resolving power ($R = 3200-4700$) during those observations.  Some additional differences in the H Balmer line strengths were found among the Be stars due to changes in their disk strengths, a common source of variability among Be stars.  We did not find any other evidence for changes in the \ion{He}{1} or other lines among any other cluster stars.  Neither the VLT nor the CTIO spectra have a high quality wavelength calibration, so we did not look for variations in radial velocity between the data sets.  

We can use our $B_\ell$ detections to test the hypothesis that the magnetic fields in NGC 3766 are fossil remnants of the original molecular cloud.  For a fossil origin, it is reasonable to assume that all of the cluster members were born with oblique dipole fields and similar magnetic fluxes but random orientations.  If we assume that magnetic flux is also conserved among these B stars (as observed by \citealt{landstreet2008}), then the present day surface field strength will be $B = B_0 (R_0/R_\star)^2$, where $B_0$ is the ZAMS surface dipole field strength.  

In the Oblique Dipole Rotator model (\citealt{stibbs1950}; \citealt{preston1967}; \citealt{auriere2007}), a magnetic dipole may be inclined relative to the rotation axis by the angle $\beta$, and the rotational axis itself may be inclined by the angle $i$.  Thus the dipole vector makes an angle $\gamma$ with the line of sight, and the net effect is the modulation of the measured $B_\ell = B_{\rm eff} \cos \gamma$ with the rotational phase, $\phi$, where:
\begin{equation}
\cos \gamma = \sin \beta \sin i \cos 2\pi (\phi - \phi_0) + \cos \beta \cos i.
\end{equation}
$B_{\rm eff}$ is an average over the observed photosphere of the star, corresponding to the observed longitudinal field when viewed along the dipole axis.  The surface dipole field strength, $B$, is related to $B_{\rm eff}$ through
\begin{equation}
B = \frac{20(3 - u)}{15+u} B_{\rm eff} \approx 3.5 B_{\rm eff}
\end{equation}
\citep{preston1967}.  Here we have used the limb darkening coefficient $u \approx 0.35$ for our sample \citep{wade1985}. 

We simulated two sets of fossil field populations with a range of magnetic fluxes $F_B = B_0 R_0^2$.  In Model A, we assigned a constant $F_B$ to all cluster stars.  Since the initial conditions at the ZAMS may have varied somewhat, Model B allows $F_B$ to vary randomly by 50\% among the cluster members (from the given value $F_B$ up to $1.5 F_B$).  
In each model, we assigned a random distribution of inclination angles, $0 \le i \le \pi$, over a spherical surface.  For the magnetic field orientation angles, we used a flat random distribution, $0 \le \beta \le \pi$, since low values of $\beta$ are roughly as probable as high values \citep{auriere2007}. 
While the dipoles may be co-aligned in such a coeval population, there is no evidence for or against assuming random orientations of $B_0$ and $B$ with our limited observations.  Since our measured $B_\ell$ are taken at random snapshots during the stars' rotation, we also allowed a random rotational phase $0 \le \phi \le 1$.  
Using 50,000 trials for each model, we determined the expected number of magnetic field detections, $N_{\rm det}$, in our sample given our detection limit $|B_\ell| > 180$ G.  The measured $N_{\rm det}$ form a Gaussian distribution with a standard deviation given by $\sigma_N$.  From the predicted detections, $B_{\rm det}$, we also determined the mean detected field, 
\begin{equation}
B_{\rm mean} = \displaystyle \frac{ \sum |B_{\rm det}| }{ N_{\rm det} },
\end{equation}
and the standard deviation of $B_{\rm mean}$, $\sigma_B$.  In every trial, we also applied a Kolmogorov-Smirnov (K-S) statistical test to determine the probability, $p$, that the observed distribution of $B_\ell$ is drawn from the simulated distributions of fossil fields.  We list the resulting $N_{\rm det}$, $\sigma_N$, $B_{\rm mean}$, $\sigma_B$, and the mean $p$ for each set of trials in Table \ref{simulations}.   

\placetable{simulations}

Based on this simple model, we tentatively rule out a fossil origin for the strong fields observed in star 170 ($F_B =$ 168,000 G~$R_\odot^2$) or even the marginal detection of star 47 ($F_B =$ 105,000 G~$R_\odot^2$).  We would expect to have detected many more magnetic B stars with much larger $B_\ell$,  and the K-S test also reveals a very low probability for such distributions.  Weaker fossil fields as observed in star 94 ($F_B =$15,000 G~$R_\odot^2$) or in star 45 ($F_B = 400$ G~$R_\odot^2$) are far more probable, but the strong fields of stars 47 and 170 are excluded from such a scenario.  These results imply that it is unlikely that the magnetic stars of NGC 3766 have a purely fossil field origin.  We caution that a fossil field might not be ruled out if fields present at the ZAMS have decayed due to non-conservative processes, not considered here.  

Non-random distributions of $\beta$, or a $\beta$ that evolves during the MS lifetime, may also occur in a coeval population with widespread fossil magnetic fields.  We investigated our models' dependence on $\beta$ by trying several alternatives to a flat, random distribution:  a distribution weighted over the surface area of a sphere, a random binary distribution ($\beta = 0$ or $\pi/2$), and even the constant  values $\beta = 0$, $\beta = \pi/4$, and $\beta = \pi/2$.  We obtained identical results will each trial, revealing that our simple assumptions about the $\beta$ distribution do not influence the validity of our models.  


\section{Conclusions}

We have detected 2 definite and 2 possible magnetic B-type stars in our survey of 15 members of the open cluster NGC 3766.  Both of our definite $B_\ell$ detections are found in stars with abnormal He abundances, and we find a marginal $B_\ell$ signature in two other cluster members that should be confirmed by higher resolution spectropolarimetric observations.  Our observations provide only a single snapshot of $B_\ell$ among the 15 stars in our survey.  Further time-resolved observations of these B-type stars are necessary to determine the intensity and topology of the fields and investigate the spatial distribution of orientation angles throughout the region.

We do not observe any trends with the detection of $B_\ell$ with $V \sin i$ or $\log g_{\rm polar}$, suggesting neither rotation nor evolution along the MS are major factors influencing the presence of magnetic fields.  If the magnetic fields are produced by a magnetic dynamo, a correlation between $B_\ell$ and $V \sin i$ would be expected.  

To examine the possibility that the detected fields have a fossil origin, we used the Oblique Dipole Rotator model to simulate populations having a uniform or slightly varying magnetic flux on the ZAMS.  Assuming conservation of the field strength with MS evolution, we compared our 4 definite and marginal detections to the modeled fossil field populations.  None of our simple fossil models successfully reproduces both our observed range in $B_\ell$ and the expected number of field detections.  Therefore we tentatively rule out a purely fossil origin for the magnetic stars in NGC 3766.

Our results are also consistent with the bimodal distribution of magnetic field strengths found by \citet{auriere2007}, who proposed a critical magnetic threshold for the stability of large scale magnetic fields among early type stars.  This or some other, unknown mechanism may contribute to the magnetic fields detected in NGC 3766.


\acknowledgments

MVM thanks the referee, Gregg Wade, for suggestions that greatly improved this manuscript.  She gratefully acknowledges Swetlana Hubrig and Marcus Sch\"oller of the European Southern Observatory for their help collecting and interpreting the data, and she appreciates helpful discussions with Stan \v{S}tefl and Thomas Rivinius about this work.  MVM is supported by an institutional grant from Lehigh University.  This work is based on observations made with ESO Telescopes at Paranal Observatory under programme ID 080.D-0383.

{\it Facilities:} \facility{VLT-UT2}

\clearpage
\begin{deluxetable}{lccccc}
\tablewidth{0pt}
\tablecaption{Journal of Observations \label{observations}}
\tablehead{
\colhead{ } &
\colhead{HJD} &
\colhead{Total exp.} &
\colhead{ } &
\colhead{$B_\ell$} &
\colhead{$\sigma$}  \\
\colhead{Star} &
\colhead{(t $-$ 2,450,000)} &
\colhead{time (s)} &
\colhead{Detection} &
\colhead{(G)} &
\colhead{(G)} }
\startdata
\phn 25  &  4550.781  &       4800  &  None      & \phn\phn       $-$96  &  61  \\ 
\phn 31  &  4549.708  &       3000  &  None      & \phn\phn\phs   12  &  59  \\
\phn 41  &  4549.762  &       4100  &  None      & \phn\phn  $-$14 &  36  \\
\phn 45  &  4550.587  &       5000  &  Marginal  & \phn       $-$185  &  53  \\
\phn 47  &  4549.533  &       1800  &  Marginal  & \phn       $-$234  &  69  \\
\phn 55  &  4549.822  &       4200  &  None      & \phn\phn    $-$46  &  37  \\
\phn 73  &  4550.525  &       1700  &  None      & \phn       $-$138  &  73  \\
\phn 83  &  4549.625  &       1640  &  None      & \phn\phn $-$55  &  36  \\
\phn 94  &  4550.845  &       4200  &  Definite  & \phn\phs      276  &  55  \\
    111  &  4550.642  &       3600  &  None      & \phn\phn\phs   54  &  33  \\
    161  &  4549.858  &       1800  &  None      & \phn\phn    $-$32  &  25  \\
    170  &  4550.708  &       4900  &  Definite  & \phs         1710  &  32  \\
    176  &  4549.888  &       1800  &  None      & \phn\phn\phn\phs  2  &  31  \\
    196  &  4549.662  &       2500  &  None      & \phn\phn $-$30  &  39  \\
    200  &  4550.885  &       2100  &  None      & \phn\phn $-$12  &  42  \\
\enddata
\end{deluxetable}

\begin{deluxetable}{lcccccccccc}
\tablewidth{0pt}
\tablecaption{Summary of Target Sample \label{targets}}
\tablehead{
\colhead{MG} &
\colhead{Webda} &
\colhead{ } &
\colhead{$V \sin i$} &
\colhead{$T_{\rm eff}$} &
\colhead{ } &
\colhead{$M_\star$ } &
\colhead{$R_\star$} &
\colhead{$R_0$} &
\colhead{ }  \\
\colhead{ID} &
\colhead{ID} &
\colhead{$y$} &
\colhead{(km s$^{-1}$)} &
\colhead{(K)} &
\colhead{$\log g_{\rm polar}$} &
\colhead{($M_\odot$) } &
\colhead{($R_\odot$) } &
\colhead{($R_\odot$) } &
\colhead{Comment} }
\startdata
\phn 25  &       291  &     10.74 &     261  &  18995  &  4.02  &  6.6   & \phn4.2  &  3.1  &  Be transient  \\ 
\phn 31  &       151  &     10.09 &     197  &  17834  &  3.99  &  6.2   & \phn4.2  &  3.0  &  Be transient  \\
\phn 41  &       130  &     10.62 & \phn 84  &  17900  &  3.84  &  6.6   & \phn5.1  &  3.1  &  He strong     \\
\phn 45  & \phn\phn8  &     10.69 & \phn 90  &  15500  &  4.61  &  3.6   & \phn1.5  &  1.5  &  Normal B star\tablenotemark{a} \\
\phn 47  & \phn   15  & \phn 8.53 &     190  &  18399  &  3.30  &  9.3   &    11.3  &  3.7  &  Be star       \\
\phn 55  & \phn   23  &     10.40 &     121  &  18000  &  3.88  &  6.6   & \phn4.9  &  3.1  &  He strong     \\
\phn 73  & \phn   26  & \phn 9.23 &     296  &  18274  &  3.49  &  8.2   & \phn8.5  &  3.5  &  Be transient  \\
\phn 83  & \phn   27  & \phn 8.34 &     111  &  18817  &  3.31  &  9.8   &    11.5  &  3.8  &  Be transient  \\
\phn 94  &       194  &     10.79 &     177  &  15650  &  3.94  &  5.0   & \phn4.0  &  2.6  &  He strong     \\
    111  & \phn   52  &     10.24 & \phn 66  &  18436  &  4.28  &  5.7   & \phn2.9  &  2.8  &  Normal B star \\
    161  & \phn   70  & \phn 9.08 & \phn 73  &  18400  &  3.49  &  8.3   & \phn8.5  &  3.5  &  Normal B star \\
    170  & \phn   94  &     10.68 & \phn 65  &  18060  &  3.82  &  6.8   & \phn5.3  &  3.1  &  He weak       \\
    176  &       212  & \phn 9.13 & \phn 54  &  21435  &  4.18  &  7.6   & \phn3.7  &  3.3  &  Normal B star \\
    196  &       239  & \phn 9.44 &     165  &  19660  &  3.87  &  7.6   & \phn5.3  &  3.3  &  Be transient  \\
    200  &       240  & \phn 9.70 &     236  &  16301  &  3.51  &  6.8   & \phn7.6  &  3.1  &  Be star       \\
\enddata
\tablenotetext{a}{Classified as a possible Be star by \citet{shobbrook1985, shobbrook1987}.  May be a transient Be star.}
\end{deluxetable}


\begin{deluxetable}{lcccccc}
\tablewidth{0pt}
\tablecaption{Simulations of Fossil Field Populations \label{simulations}}
\tablehead{
\colhead{ } &
\colhead{$F_B$ } &
\colhead{ } & 
\colhead{ } & 
\colhead{$B_{mean}$} &
\colhead{$\sigma_B$} &
\colhead{Mean $p$} \\
\colhead{Model} &
\colhead{(G~$R_\odot^2$) } &
\colhead{$N_{det}$} &
\colhead{$\sigma_N$} &
\colhead{(G)} & 
\colhead{(G)} &
\colhead{(\%)} }
\startdata
Model A: 
         &        170,000  &     12.8  &  1.3  &     1965  &     557  &  10.3  \\
         &        100,000  &     11.0  &  1.4  &     1315  &     383  &  22.8  \\
         & \phn    80,000  &     10.1  &  1.4  &   1138  &    335  &  31.4  \\
         & \phn    60,000  & \phn 9.0  &  1.4  & \phn 935  & 283  &  45.1  \\
         & \phn    40,000  & \phn 7.3  &  1.4  & \phn 742  & 239  &  66.2  \\
         & \phn    30,000  & \phn 6.2  &  1.5  & \phn 629  & 218  &  80.0  \\
         & \phn    20,000  & \phn 4.3  &  1.4  & \phn 549  & 235  &  95.7  \\
         & \phn    15,000  & \phn 2.8  &  1.1  & \phn 550  & 279  &  99.3  \\
         & \phn    10,000  & \phn 1.5  &  0.7  & \phn 525  & 302  &  99.9  \\
         & \phn \phn 5,000 & \phn 0.6 & 0.3  & \phn 291  & 215  &  99.9  \\
\\
Model B: 
         &        170,000  &     13.2  &  1.3  &     2373  &     683  & \phn 7.9  \\
         &        100,000  &     11.8  &  1.4  &     1547  &     456  & 16.5  \\
         & \phn    80,000  &     11.0  &  1.4  &   1322  &     394  & 23.2  \\
         & \phn    60,000  & \phn 9.8  &  1.4  & 1089  &     331  & 34.5  \\
         & \phn    40,000  & \phn 8.2  &  1.4  & \phn 843  & 267  & 55.2  \\
         & \phn    30,000  & \phn 7.0  &  1.4  & \phn 717  & 240  & 69.6  \\
         & \phn    20,000  & \phn 5.4  &  1.5  & \phn 585  & 225  & 88.7  \\
         & \phn    15,000  & \phn 3.8  &  1.3  & \phn 554  & 253  & 97.0  \\
         & \phn    10,000  & \phn 2.1  &  1.0  & \phn 542  & 302  & 99.8  \\
         & \phn\phn 5,000 & \phn 0.9  &  0.6  & \phn 367  & 246  & 99.9 \\
\enddata
\end{deluxetable}

\clearpage

\begin{thebibliography}

\bibitem[Alecian et al.(2008)]{alecian2008} 
Alecian, E., et al.  2008, \aap, 481, L99

\bibitem[Auri\`ere et al.(2007)]{auriere2007} 
Auri\`ere, M., et al.  2007, \aap, 475, 1053

\bibitem[Bagnulo et al.(2006)]{bagnulo2006} 
Bagnulo, S., Landstreet, J.~D., Mason, E., Andretta, V., Silaj, J., \& Wade, G.~A.  2006, \aap, 450, 777

\bibitem[Bagnulo et al.(2002)]{bagnulo2002} 
Bagnulo, S., Szeifert, T., Wade, G.~A., Landstreet, J.~D., \& Mathys, G. 2002, \aap, 389, 191 

\bibitem[Casini \& Landi degl'Innocenti(1994)]{casini1994} 
Casini, R., \& Landi degl'Innocenti, E.\ 1994, \aap, 291, 668

\bibitem[Charbonneau \& MacGregor(2001)]{charbonneau2001} 
Charbonneau, P., \& MacGregor, K.~B.  2001, \apj, 559, 1094

\bibitem[Huang \& Gies(2006)]{huang2006} 
Huang, W., \& Gies, D.~R.  2006, \apj, 648, 591

\bibitem[Landstreet(1982)]{landstreet1982} 
Landstreet, J.~D. 1982, \apj, 258, 639 

\bibitem[Landstreet et al.(2008)]{landstreet2008} 
Landstreet, J.~D., et al.  2008, \aap, 481, 465 

\bibitem[MacDonald \& Mullan(2004)]{macdonald2004} 
MacDonald, J., \& Mullan, D.~J.  2004, \mnras, 348, 702

\bibitem[MacGregor \& Cassinelli(2003)]{macgregor2003} 
MacGregor, K.~B., \& Cassinelli, J.~P.  2003, \apj, 586, 480 

\bibitem[Mathys(1988)]{mathys1988}
Mathys, G. 1988, \aap, 189, 179

\bibitem[McSwain \& Gies(2005)]{mcswain2005b} 
McSwain, M.~V., \& Gies, D.~R. 2005, \apjs, 161, 118

\bibitem[McSwain et al.(2008)]{mcswain2008} 
McSwain, M.~V., Huang, W., Gies, D.~R., Grundstrom, E.~D., \& Townsend, R.~H.~D.  2008, \apj, 672, 590

\bibitem[Moitinho et al.(1997)]{moitinho1997}
Moitinho, A., Alfaro, E.~J., Yun, J.~L., \& Phelps, R.~L. 1997, \aj, 113, 1359

\bibitem[Neiner(2007)]{neiner2007} 
Neiner, C.  2007, Active OB-Stars: Laboratories for Stellar and Circumstellar Physics, 361, 91 

\bibitem[Petit et al.(2008)]{petit2008} 
Petit, V., Wade, G.~A., Drissen, L., Montmerle, T., \& Alecian, E.\ 2008, MNRAS, 387, L23

\bibitem[Preston(1967)]{preston1967}
Preston, G.\ W.  1967, \apj, 150, 547

\bibitem[Schaller et al.(1992)]{schaller1992}
Schaller, G., Schaerer, D., Meynet, G., \& Maeder, A. 1992, \aaps, 96, 269

\bibitem[Shobbrook(1985)]{shobbrook1985}
Shobbrook, R. R. 1985, \mnras, 212, 591
         
\bibitem[Shobbrook(1987)]{shobbrook1987}
Shobbrook, R. R. 1987, \mnras, 225, 999

\bibitem[Spruit(2002)]{spruit2002} 
Spruit, H.~C.\ 2002, \aap, 381, 923

\bibitem[Stibbs(1950)]{stibbs1950}
Stibbs, D.\ W.\ N. 1950, \mnras, 110, 395

\bibitem[Tadross(2001)]{tadross2001} 
Tadross, A.~L.\ 2001, New Astronomy, 6, 293

\bibitem[Wade \& Rucinski(1985)]{wade1985} 
Wade, R.~A., \& Rucinski, S.~M. 1985, \aaps, 60, 471 

\end{thebibliography}
\end{document}